\documentclass{article}
\usepackage{spconf,amsmath,graphicx}
\usepackage{amssymb}
\usepackage{subcaption}
\usepackage{siunitx}
\usepackage{stfloats}
\usepackage{enumitem}
\usepackage{xparse,mathtools}
\usepackage{pgfplots}
\usepackage{tikz}
\usepackage{graphicx}

\newcommand\mycopyrighttext{%
	\footnotesize © 2023 IEEE. Personal use of this material is permitted. Permission from IEEE must be obtained for all other uses, in any current or future media, including reprinting/republishing this material for advertising or promotional purposes, creating new collective works, for resale or redistribution to servers or lists, or reuse of any copyrighted component of this work in other works.}
\newcommand\mycopyrightnotice{%
	\begin{tikzpicture}[remember picture,overlay]
		\node[anchor=south,yshift=10pt] at (current page.south) {\fbox{\parbox{\dimexpr\textwidth-\fboxsep-\fboxrule\relax}{\mycopyrighttext}}};
	\end{tikzpicture}%
}

\title{Estimating and analyzing neural information flow using signal processing on graphs}
\name{Felix Schwock$^{1}$\sthanks{corresponding author: fschwock@uw.edu}, Julien Bloch$^2$, Les Atlas$^1$, Shima Abadi$^1$, Azadeh Yazdan-Shahmorad$^{1,2}$\sthanks{This research was funded by the American Heart Association (FS, AY), the National Institute of Health R01 NS119593 (JB, AY) and R01 MH125429 (FS, AY), and the Washington Research Foundation (AY).}}
\address{$^1$Department of Electrical and Computer Engineering University of Washington, Seattle \\
$^2$Department of Bioengineering, University of Washington, Seattle \\
}
\begin{document}
\ninept


\maketitle
%

\mycopyrightnotice

\begin{abstract}
Correlating neural communication in brain networks with behavior and cognition can provide fundamental insights into the functionality of both healthy and diseased brains.
We demonstrate how communication in the brain can be estimated from recorded neural activity using concepts from graph signal processing.
The communication is modeled as a flow signals on the edges of a graph and naturally arises from a graph diffusion process.
We apply the diffusion model to micro-electrocorticography (ECoG) recordings from sensorimotor cortex of two non-human primates to estimate the neural communication flow during excitatory optogenetics.
Comparisons with a baseline model demonstrate that adding the neural flow can improve ECoG predictions.
Finally, we demonstrate how the neural flow can be decomposed into a gradient and rotational component and show that the gradient component depends on the location of stimulation.
This technique, for the first time, offers the opportunity to study neural communication on an unprecedented spatiotemporal scale.
\end{abstract}
\begin{keywords}
network neuroscience, neural communication, graph diffusion, edge signals, non-human primates, optogenetics
\end{keywords}
\section{Introduction}
\label{sec:intro}

A deeper understanding of the internal communication processes in the brain may be crucial for developing more effective treatments for many neurological diseases \cite{avena-koenigsberger_2018, voytek_2015}.
Furthermore, with the advancement of electrode arrays that expand across multiple brain areas and provide simultaneous recordings from hundreds of channels, developing novel tools to analyze and model the dynamics of brain neural networks is much needed.
Recently, graphs have emerged as promising tools to analyze neural processes.
Thereby most research has focused on studying properties of brain graphs that were derived from structural or functional measurements of the brain \cite{bullmore_2009, bassett_2017}.
More recently, graph signal processing (GSP) has been proposed as a way to study neural signals that are observed at the nodes of an underlying brain network \cite{huang_2016, huang_2018, mortaheb_2019}.
The central idea of GSP is that, rather than focusing on the graph itself, we are interested in processing signals indexed by the nodes of the graph, for example, by finding a graph spectral representation of the signal, or performing filtering in the graph domain \cite{shuman_2013}.
By focusing on graph signals, this framework therefore offers a principle way to study dynamic network and communication processes in the brain.

GSP in neuroscience has, to the best of our knowledge, exclusively focused on signals defined on the nodes of an underlying brain graph.
Our goal is to shift the focus from signals at the nodes to flow signals residing on the edges of a graph that characterize the information flow in the brain. 
Specifically, we will address the following questions:
(1) How can we leverage methods from graph signal processing to estimate the neural flow in the brain?
(2) How can we naturally analyze and process these flow signals?

This new perspective allows for modeling neural communication on the order of milliseconds rather than seconds or minutes as achieved by other techniques in neuroscience such as Granger causality \cite{granger_1969, dhamala_2008, seth_2015}.
To test our theoretical findings, we will apply our model to estimate neural flow from electrophysiological recordings obtained from monkeys during excitatory optogenetics.


The remainder of the paper is outlined as follows:
In Sec.~\ref{sec:diff_model} we describe the model that turns neural recordings into a flow signal.
In Sec.~\ref{sec:est_neural_flow_monkey} the model is used to estimate the neural flow in two non-human primates during a stimulation experiment.
Sec.~\ref{sec:analyze_neural_flow} describes ways to analyze the neural flow signal by decomposing it into different components.
Finally, Sec.~\ref{sec:conclusion} summarizes the main findings and points towards future research directions.

\section{Estimating Neural Flows from Time Series of Neural Activity}\label{sec:diff_model}

In this section, we describe how a graph diffusion model can be used to estimate neural flow signals from vector time series of neural activity.
Diffusion models have been used in the GSP community to describe the dynamic behavior of network data \cite{shuman_2013, grassi_2018, thanou_2017}, as well as in the neuroscience community to model functional connectivity \cite{abdelnour_2014}.
Instead of focusing on functional connectivity, here we demonstrate how neural activity on the nodes of a graph can be transformed into an edge flow signal using a diffusion process.

\begin{figure}[tb!]
	\centerline{\includegraphics[width=0.45\textwidth]{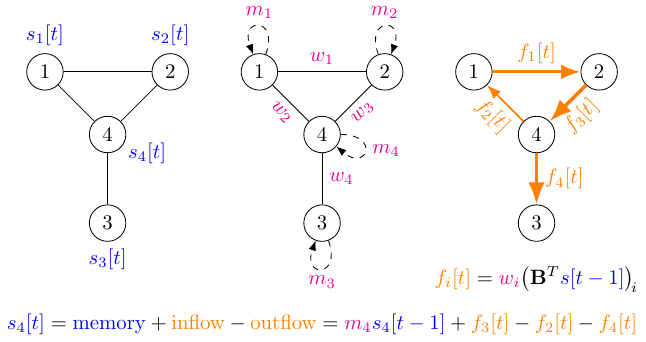}}

	\vspace{-0.2cm}
	\caption{Illustration of diffusion model. The three graphs show from left to right the node signal, the model parameters, and the edge flow, respectively. The equation at the bottom illustrates the diffusion model \eqref{eq:diff_flow_model} for Node~4.}
	\label{fig:diffusion_model}
	
\end{figure}

We start by assuming a graph with $N$ nodes and $E$ edges.
Furthermore, time series of neural activity are measured at the nodes of the graph.
We will use $s[t] \in \mathbb{R}^{N \times 1}$ to denote the neural activity at time $t$ across all nodes.
An example of this is shown on the left of Fig.~\ref{fig:diffusion_model}, with the graph topology (nodes and edges) in black and the node time series $s_i[t]$ in blue.
Our goal is to use the given graph topology and observed node time series to estimate the time dependent information flow along the edges of the graph.
This flow is illustrated on the right in Fig.~\ref{fig:diffusion_model}.

\subsection{$\mathbf{1^\mathrm{st}}$ Order Diffusion Model}

To estimate the flow, we use a parameterized diffusion model that naturally gives rise to an edge flow and whose parameters can be estimated from the observed node time series.
In a nutshell, for each node the model expresses the node signal at the current time step as its weighted own past (memory) plus the inflow minus the outflow from all neighboring nodes.
To describe this mathematically, we first encode the graph topology using the node-to-edge incidence matrix $\mathbf{B} \in \mathbb{R}^{N \times E}$, where each column represents an edge $e = (i, j)$ with tail node $i$ and head note $j$.
$\mathbf{B}$ is defined such that $B_{ie} = -B_{je} = -1$ and $B_{ke} = 0$ otherwise.
For every edge, it is thereby arbitrary which of the incident nodes is the tail and head node.
For more details and examples for $\mathbf{B}$, we refer the reader to \cite{schaub_2021}.
Using $\mathbf{B}$, our diffusion model can be written as
\begin{equation}\label{eq:diff_flow_model}
	s[t] = \left(\mathbf{M} - \mathbf{B}\mathbf{W}\mathbf{B}^T\right)s[t-1],
\end{equation}
where $\mathbf{M} = \text{diag}\left(\mathbf{m}\right) \in \mathbb{R}^{N \times N}$ and $\mathbf{W} = \text{diag}\left(\mathbf{w}\right) \in \mathbb{R}^{E \times E}$ are diagonal matrices containing the node and edge parameters, respectively (Fig.~\ref{fig:diffusion_model} middle plot).
To better understand the flow component of the model, we take a closer look at the $\mathbf{B}\mathbf{W}\mathbf{B}^T s[t-1]$ term and assign a physical meaning to each operation:

\begin{itemize}[noitemsep, topsep=0pt]
	\item $\mathbf{B}^T s[t-1]$: $\mathbf{B}^T$ acts as a discrete gradient operator. If $s[t]$ represents a local field potential time series (i.e., voltages), $\mathbf{B}^T s[t-1]$ computes the voltage gradient for each edge.
	\item $f[t] = \mathbf{W}\mathbf{B}^T s[t-1]$: Multiplying each gradient by the corresponding edge weight $w_i$ yields the flow.
	In analogy to resistive circuits, $w_i$ can be interpreted as the conductivity between two nodes, so that conductivity times potential gradient yields the current.
	\item $\mathbf{B}f[t]$: $\mathbf{B}$ acts as the discrete divergence operator.
	That is, for each node we compute the net flow, which is the sum of all inflows minus the sum of all outflows.
\end{itemize}

\subsection{$\mathbf{K^\mathrm{th}}$ Order Diffusion Model}

A limitation of the model in \eqref{eq:diff_flow_model} is that only the neural activity at time $t-1$ is used to predict the neural activity at time $t$, but no prior time steps are incorporated.
This means that all edges in the graph encode a time delay of $t=1$, i.e., the individual weights can be expressed as $m_i \delta[t-1]$ and $w_i \delta[t-1]$.
However, it is reasonable to assume that the flow between different edges experiences different time delays.
To allow for that, we can replace the node and edge weights by linear filters $m_i[t]$ and $w_i[t]$ of arbitrary order $K$
\begin{align}\label{eq:edge_weights_as_filters}
	\begin{split}
		m_i[t] &= m_{i,1}\delta[t-1] + ... + m_{i,K}\delta[t-K] \\
		w_i[t] &= w_{i,1}\delta[t-1] + ... + w_{i,K}\delta[t-K].
	\end{split}
\end{align}
The resulting $K^\mathrm{th}$ order diffusion model is given by
\begin{equation}\label{eq:diff_flow_model_generalized}
	s[t] = \sum_{i=1}^{K}\left(\mathbf{M}_k - \mathbf{B}\mathbf{W}_k\mathbf{B}^T\right)s[t-k],
\end{equation}
where $\mathbf{M}_k = \text{diag}\left(\mathbf{m}_k\right) \in \mathbb{R}^{N \times N}$ and $\mathbf{W}_k = \text{diag}\left(\mathbf{w}_k\right) \in \mathbb{R}^{E \times E}$ are diagonal matrices containing the node and edge parameters for the $k^\mathrm{th}$ lag.
The node signals can now be transformed into the edge flow domain via
\begin{equation}\label{eq:node_signal_to_flow}
	f[t] = \sum_{k=1}^{K}\mathbf{W}_k\mathbf{B}^T s[t-k].
\end{equation}


%
%

%
%

\subsection{Estimating the Diffusion Model Parameters}

The parameters of the $K^\mathrm{th}$ order diffusion model $\mathbf{M}_k$ and $\mathbf{W}_k$ can be estimated from neural recordings by minimizing the least squares prediction error
\begin{equation}\label{eq:pred_error}
	e = \sum_{t \in T_e} \vert\vert \hat{s}[t] - s[t] \vert\vert_2^2,
\end{equation}
where $\hat{s}[t]$ is the predicted neural activity according to \eqref{eq:diff_flow_model_generalized} and $s[t]$ the observed neural activity.
$T_e$ denotes the set of all time points $t$ used for fitting the model.
Note that this set does not need to contain successive indices.
It can be shown that minimizing $e$ w.r.t. $\mathbf{M}_k$ and $\mathbf{W}_k$ can be formulated as solving a quadratic program \cite{boyd_vandenberghe_2004}.
No regularizers on the parameters have been used in this work.

\section{Estimating Neural Flow in two non-human primates}\label{sec:est_neural_flow_monkey}

The diffusion model \eqref{eq:diff_flow_model_generalized} was applied to neural recordings from two rhesus macaque monkeys during a stimulation experiment for various model orders $K$.
The neural activity was recorded in the form of local field potential (LFP) measurements by a 96 electrode micro-electrocorticography ($\mathrm{\mu}$-ECoG) array placed over the primary somatosensory and motor cortex.
Stimulation of excitatory neurons was enabled through an optogenetic interface \cite{yazdan-shahmorad_2015, yazdan-shahmorad_2016, khateeb_2019}.
During the experiment, short laser pulses by one or two lasers were used to repeatedly stimulate the brain at a rate of \num{5} or \SI{7}{\hertz} for \SI{10}{\minute} blocks at specific locations that varied between the 63 experimental sessions.
Here we will refer to each stimulation incident by either one or two lasers as a \textit{trial}.
Thus, every stimulation block consists of \num{3000} or \num{4200} trials depending on whether the stimulation rate is \num{5} or \SI{7}{\hertz}.
If two lasers were used (53 out of 63 sessions), stimulation between the two lasers was delayed by \num{10}, \num{30}, \num{70}, or \SI{100}{\milli\second} depending on the session.
Each experimental session consists of 5 stimulation blocks interrupted by resting blocks during which no stimulation was performed.
LFPs are recorded with a sampling frequency of \SI{1017.25}{\hertz}.
More details on the experimental setup and data collection can be found in \cite{yazdan_2018, bloch_2022}.

\begin{figure*}[htb!]
	\centering
	\begin{minipage}[b]{.37\linewidth}
		\centering
		\includegraphics[width=\textwidth]{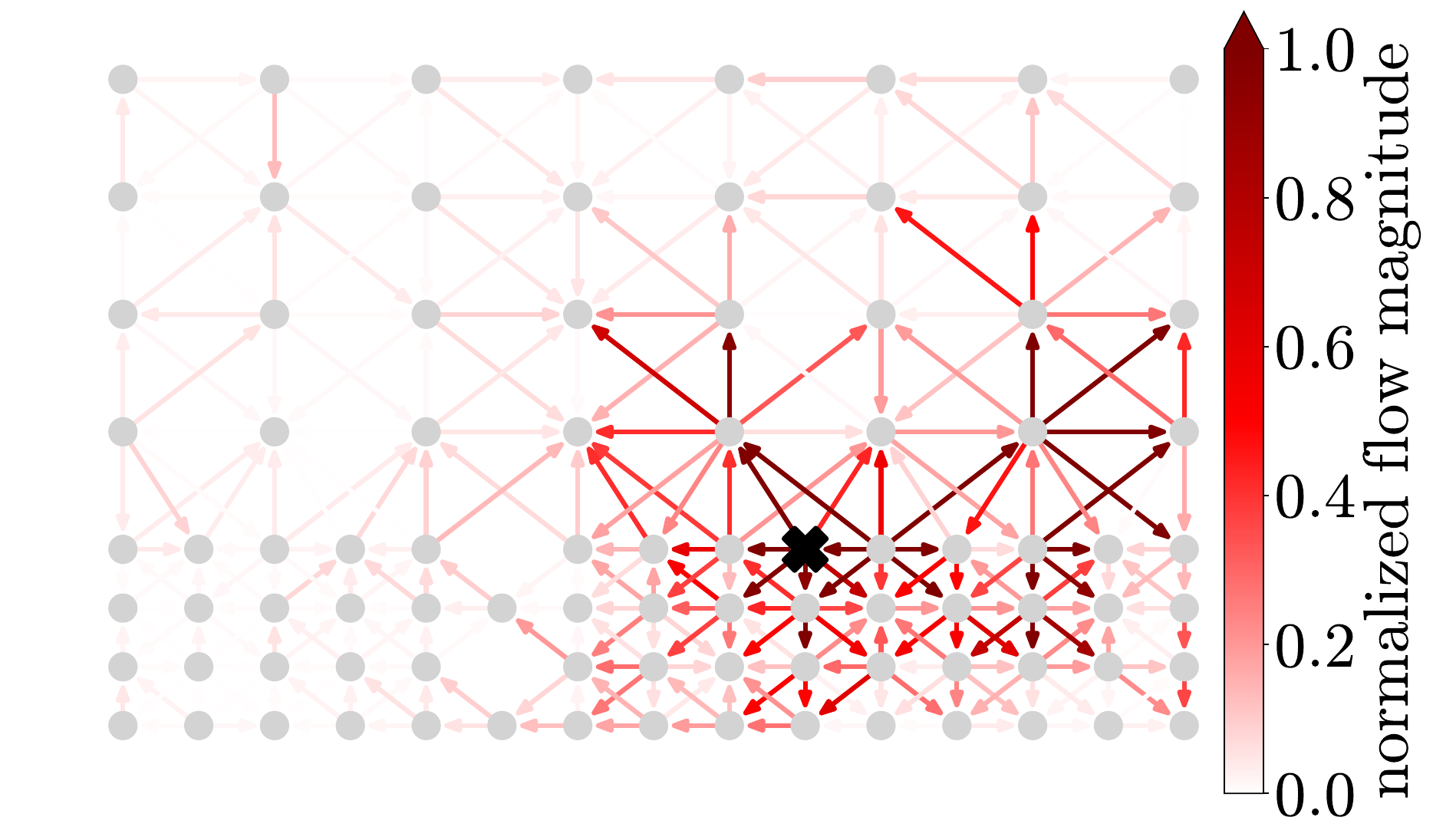}
		\vspace{-0.7cm}
		
		\centerline{(a) \SI{5}{\milli\second} after stimulation}\medskip
	\end{minipage}
	~
	\begin{minipage}[b]{.37\linewidth}
		\centering
		\includegraphics[width=\textwidth]{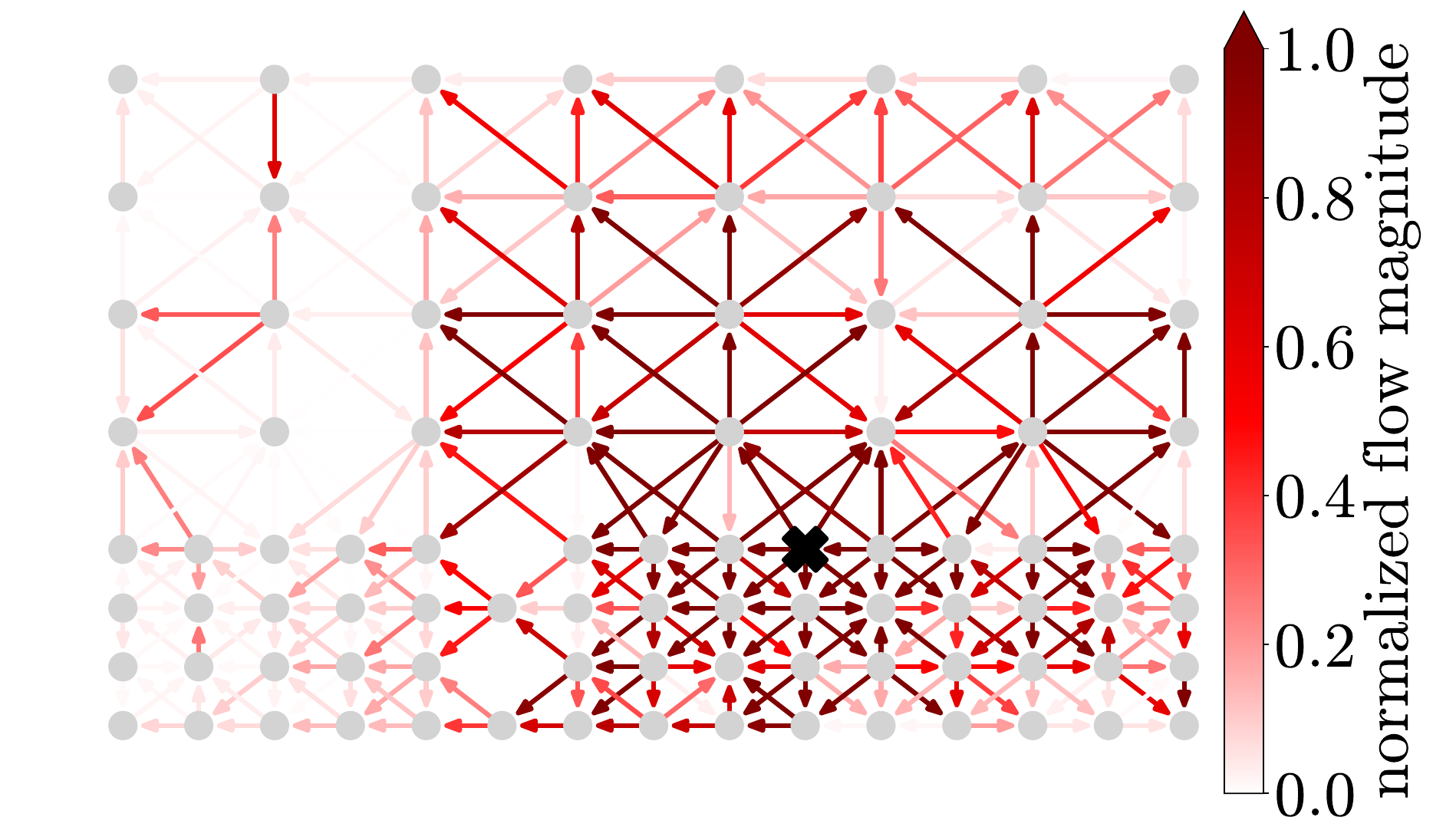}
		\vspace{-0.7cm}
		
		\centerline{(b) \SI{10}{\milli\second} after stimulation}\medskip
	\end{minipage}
	\vspace{-0.3cm}
	\captionof{figure}{Normalized neural flow of a single experimental session 5 and \SI{10}{\milli\second} after stimulation. The location of stimulation is indicated by the black cross. The flow was estimated using a $5^\mathrm{th}$ order diffusion model (Eq.~\eqref{eq:diff_flow_model_generalized}).}\label{fig:neural_flow}
	
	\vspace{-0.3cm}
\end{figure*}

%

To describe the neural information flow induced by the laser stimulation, first the mapping of the $\mathrm{\mu}$-ECoG array is used to construct a sparsely connected graph, where each node is connected approximately to its 8 nearest neighbors.
The LFP time series measured by each electrode are demeaned and treated as the node signals $s[t]$.
The defined graph topology along with $s[t]$ starting at the time of stimulation of the first laser until \SI{30}{\milli\second} after stimulation from the second laser (first laser if only one is used) are used to estimate the diffusion model parameters $\mathbf{M}_k$ and $\mathbf{W}_k$.
The flow can now be computed according to \eqref{eq:node_signal_to_flow}.
Examples of neural flow for \SI{5}{\milli\second} and \SI{10}{\milli\second} after stimulation are shown in Fig.~\ref{fig:neural_flow} for a model order of $K=5$.
The location of stimulation is indicated by the cross.
One can see that \SI{5}{\milli\second} after stimulation the flow magnitude is mainly located near the stimulation site, whereas at \SI{10}{\milli\second} the flow has spread further into the network.


To asses the importance of the flow part of the model, we demonstrate that the estimated flow can improve LFP predictions over a \textit{no-flow} baseline model.
The \textit{no-flow} model is obtained by using \eqref{eq:diff_flow_model_generalized} with the constraint $\mathbf{W}_k = 0 \; \forall k$. 
That is the \textit{no-flow} model reduces to a simple $K^\mathrm{th}$ order autoregressive model with no connections between the nodes.
Both the \textit{flow} and \textit{no-flow} model are fitted to all 63 sessions using \SI{80}{\percent} of the trails (training set) in each session.
Then, for each session, the estimated model parameters are used for a one-step-ahead prediction on the remaining \SI{20}{\percent} (test set) and the root-mean squared error (RMSE) between the measured LFPs $s[t]$ and predicted LFPs $\hat{s}[t]$ is computed for each time step $t$ in the test set.
For each stimulation block, the model is independently fitted to the first \SI{80}{\percent} of the trials and evaluated on the remaining \SI{20}{\percent}.
Finally, we compute by how many percent the \textit{flow} model improves the LFP predictions over the \textit{no-flow} model for each session $s$ as 
\begin{equation}\label{eq:pred_improve}
	I(s) = 100 \cdot \underset{t}{\mathrm{median}}\left\lbrace \frac{RMSE_{s,\text{no flow}}[t] - RMSE_{s,\text{flow}}[t]}{RMSE_{s,\text{no flow}}[t]} \right\rbrace.
\end{equation}

\begin{figure}[tb]
	\centering
	\begin{minipage}[b]{0.48\linewidth}
		\centering
		\includegraphics[width=\textwidth]{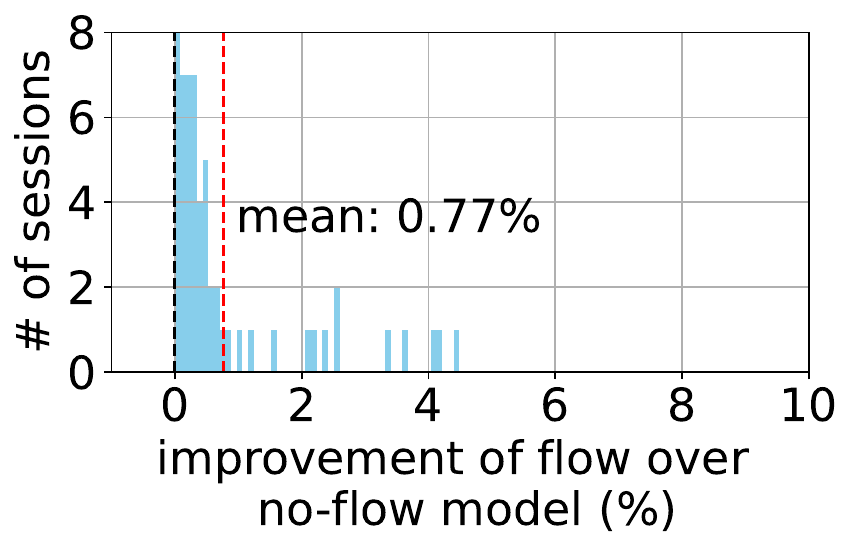}
		
		\centerline{(a) $K=1$}\medskip
	\end{minipage}
	~
	\begin{minipage}[b]{0.48\linewidth}
		\centering
		\includegraphics[width=\textwidth]{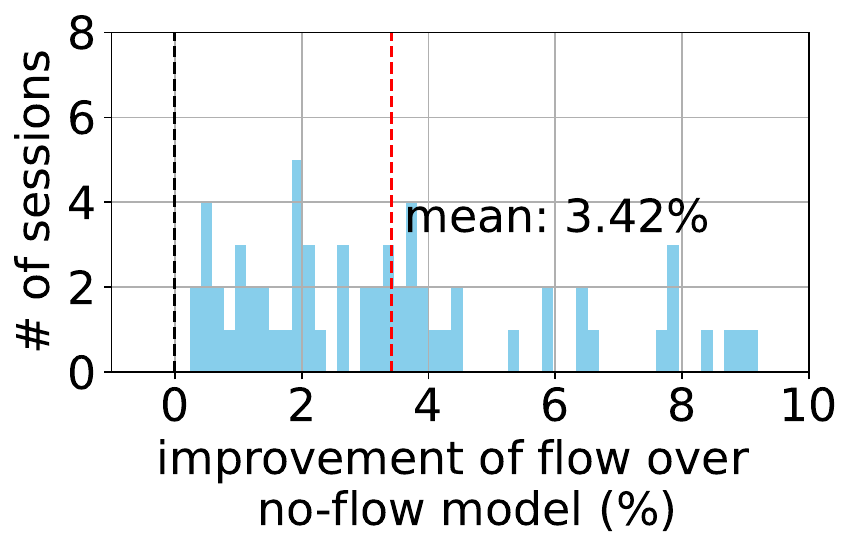}
		
		\centerline{(c) $K=9$}\medskip
	\end{minipage}
	
		
	
	\vspace{-0.3cm}
	\captionof{figure}{Histogram of the improvement in LFP predictions of \textit{flow} over \textit{no-flow} model for 63 experimental sessions and two differnt model orders according to \eqref{eq:pred_improve}. The black dashed line indicates \SI{0}{\percent} improvement. The red dashed lines show the average improvement across all sessions. For all 63 sessions, our proposed \textit{flow} model performs better than the \textit{no-flow} model for $K = 9$.}\label{fig:prediction_performance}
\end{figure}

\noindent The distribution of $I(s)$ for all 63 sessions is shown in Fig.~\ref{fig:prediction_performance} for $K=1$, and \num{9}.
It is notable that, for $K=9$, the \textit{flow} model always predicts the LFPs better than the \textit{no-flow} model.
This is in fact true for all model orders grater than \num{1} (for $K=1$ the \textit{flow} model predicts the LFPs better than the \textit{no-flow} model for \num{62} out of \num{63} sessions).
With increasing model order, the average improvement across all sessions increases from \SI{0.77}{\percent} for $K=1$ (statistical significant; p-value: $2.2\cdot10^{-4}$; Student's t-test) to \SI{3.42}{\percent} for $K=9$ (statistical significant; p-value: $2.3\cdot10^{-12}$; Student's t-test).

\section{Analyzing Neural Flow Signals}\label{sec:analyze_neural_flow}

So far we have illustrated how neural flow arises as the result of a graph diffusion process.
In this section, we will show some preliminary results demonstrating how we can analyze the obtained flow signals and how they depend on the location of laser stimulation.

\subsection{Decomposing Flow Into Gradient and Rotational Component}

Only very recently, efforts have been made to develop a signal processing theory, for signals supported on edges and higher order networks \cite{schaub_2021, barbarossa_2020, schaub_segarra_2018}.
This includes defining a spectral representation based on the Hodge-Laplacian that can be used to decompose a flow signal into different gradient and rotational modes, as well as defining flow filters based on this notion \cite{yang_2021}.
Here we will demonstrate how the neural flow can be decomposed into its gradient and rotational component.

Any flow signal defined on the edges of a graph can be decomposed into a gradient and rotational component \cite{schaub_segarra_2018}:
\begin{equation}\label{eq:flow_decomposition}
	f = f_\mathrm{grad} + f_\mathrm{rot},
\end{equation}
where $f_\mathrm{grad}$ and $f_\mathrm{rot}$ have the following properties:
\begin{align}\label{eq:flow_divergence}
	\mathbf{B} f_\mathrm{grad} & > 0 \\
	\mathbf{B} f_\mathrm{rot} & = 0.
\end{align}
Recall that $\mathbf{B}$ applied to a flow signal computes its divergence, i.e., the sum of all inflow minus outflow at each node.
That is, the rotational component $f_\mathrm{rot}$ is flow preserving at each node, thus describing circular flows, whereas the gradient flow (non-circular flow) has non-zero divergence for some of the nodes causing those nodes to act more as sources (outflow $>$ inflow) or sinks (inflow $>$ outflow).

The gradient flow can be further decomposed into $N-1$ orthogonal modes
\begin{align}\label{eq:gradeint_decomposition}
	\begin{split}
		f_\mathrm{grad} = & f_\mathrm{grad,1} + ... +  f_{\mathrm{grad},N-1} \\
		& \mathrm{with} \; f_{\mathrm{grad},i} \perp f_{\mathrm{grad},j} \; \mathrm{for} \; i \neq j,
	\end{split}
\end{align}
where $N$ is the number of nodes in the graph.
These gradient flow modes can be ordered in increasing amount of divergence.
That is
\begin{equation}\label{eq:gradeint_decomposition_divergence}
	\mathbf{B} \frac{f_\mathrm{grad,1}}{\vert\vert f_\mathrm{grad,1} \vert\vert_2} \leq \mathbf{B} \frac{f_\mathrm{grad,2}}{\vert\vert f_\mathrm{grad,2} \vert\vert_2} \leq ... \leq \mathbf{B} \frac{f_{\mathrm{grad},N-1}}{\vert\vert f_{\mathrm{grad},N-1} \vert\vert_2}.
\end{equation}
Gradient modes with small divergence can be interpreted as \textit{smooth} or low-frequency gradient flows on the graph, whereas modes with high divergence are said to be high-frequency gradient flows.
Based on this notion, we can filter out the high-frequency gradient modes to obtain a smooth gradient flow \cite{yang_2021}.

\begin{figure}[tb!]
	\centerline{\includegraphics[width=0.45\textwidth]{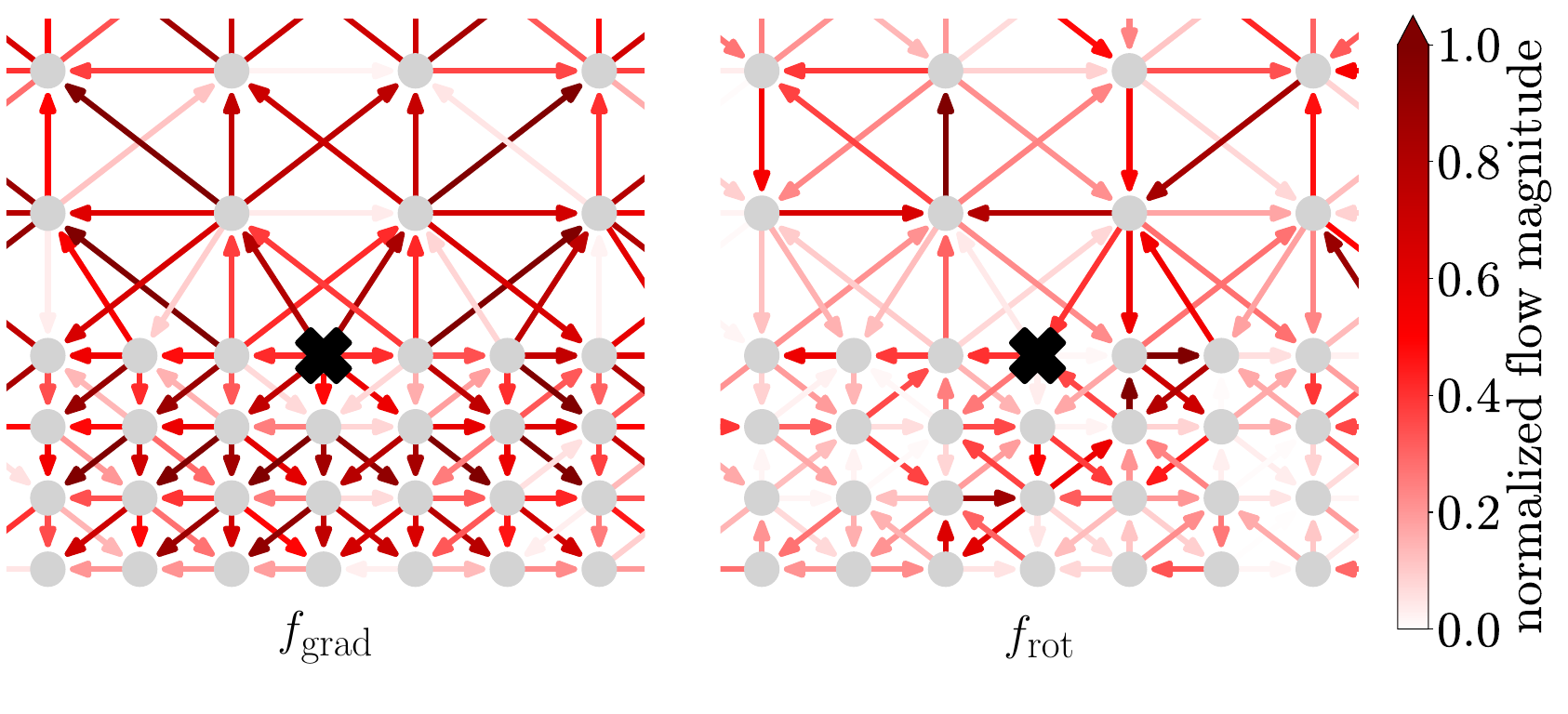}}

	\vspace{-0.2cm}
	\captionof{figure}{Zoomed-in gradient flow $f_\mathrm{grad}$ and rotational flow $f_\mathrm{rot}$ for a single experimental session \SI{10}{\milli\second} after stimulation around the location of stimulation (black cross). A model of oder $K=5$ is was used to estimate the flow.}\label{fig:neural_flow_decomposition}
	
\end{figure}

The decomposition of the neural flow signal $f$ \SI{10}{\milli\second} after stimulation into $f_\mathrm{grad}$ and $f_\mathrm{rot}$ for the same experimental session as in Fig.~\ref{fig:neural_flow} and model order $K=5$ is illustrated in Fig.~\ref{fig:neural_flow_decomposition}.
For a better visual representation, we have filtered out the gradient flow modes with high divergence (\num{3}~dB cutoff approximately at $f_\mathrm{grad,14}$) and only plotted the filtered (smooth) gradient flow in Fig.~\ref{fig:neural_flow_decomposition} (left).
We expect that the neural activity spreads from the stimulation location into other parts of the network.
This is in agreement with the gradient flow signal in Fig.~\ref{fig:neural_flow_decomposition} showing flow spreading away from the stimulation site (black cross).

%
%

The behavior of the rotational flow illustrated in Fig.~\ref{fig:neural_flow_decomposition} (right) is less intuitive and a decomposition into different modes is less straight forward.
Nevertheless, the rotational flow may be of value as it cannot be related directly to a signal on the nodes of the graph, but instead can only be studied in the edge flow domain\footnote{On the other hand, the gradient flow can be turned into a node signal by taking its divergence. This node signal is uniquely related to the corresponding gradient flow up to an additive constant.}.
In the future, we plan to analyze the rotational flow and its dependence on the location of stimulation and other experimental parameters.

\subsection{Relation Between Gradient Flow and Stimulation Location}

Fig.~\ref{fig:neural_flow_decomposition}(a) indicates that the gradient flow depends on the location of stimulation.
To quantify this, we can use the gradient flow to determine the location of the \textit{global broadcaster} $x_\mathrm{b}$ in the network and compare it to the stimulation location $x_\mathrm{s}$.
We hypothesize that $x_\mathrm{b}$ is significantly closer to $x_\mathrm{s}$ than a randomly placed broadcaster.
The steps to compute the location of the global broadcaster are as follows:
\begin{enumerate}[topsep=0pt,itemsep=-1ex,partopsep=1ex,parsep=1ex]
	\item Treat the entries of the gradient flow component as edge weights of a directed graph and construct the adjacency matrix $A$ of this directed graph. If we denote the entry of $f_\mathrm{grad}$ corresponding to the edge between node $i$ and $j$ as $f_{\mathrm{grad}}^{\lbrace i,j \rbrace}$, $A(i,j) = f_{\mathrm{grad}}^{\lbrace i,j \rbrace}$ if there is a positive flow from node $i$ to node $j$, $A(j,i) = -f_{\mathrm{grad}}^{\lbrace i,j \rbrace}$ if there is a positive flow from $j$ to $i$, and $A(i,j) = 0$ else.
	\item Compute the aggregate downstream reachability (ADR) for each node according to \cite{delacruzcabrera_2019}. For a given node $n$, the ADR denotes how well any other node in the network can be reached from $n$ following the direction of the flow. That is, a high ADR for a node indicates flow spreading away from this node.
	\item Using the physical location of the nodes $l(n)$ in the recording array, compute the center of mass of the ADR distribution across the nodes and denote this as the location of the global broadcaster:
	\begin{equation}\label{eq:center of mass}
		x_b = \sum_{n=1}^{N} l(n) \cdot \mathrm{ADR}(n).
	\end{equation}
\end{enumerate}

\begin{figure}[tb!]
	\centerline{\includegraphics[width=0.3\textwidth]{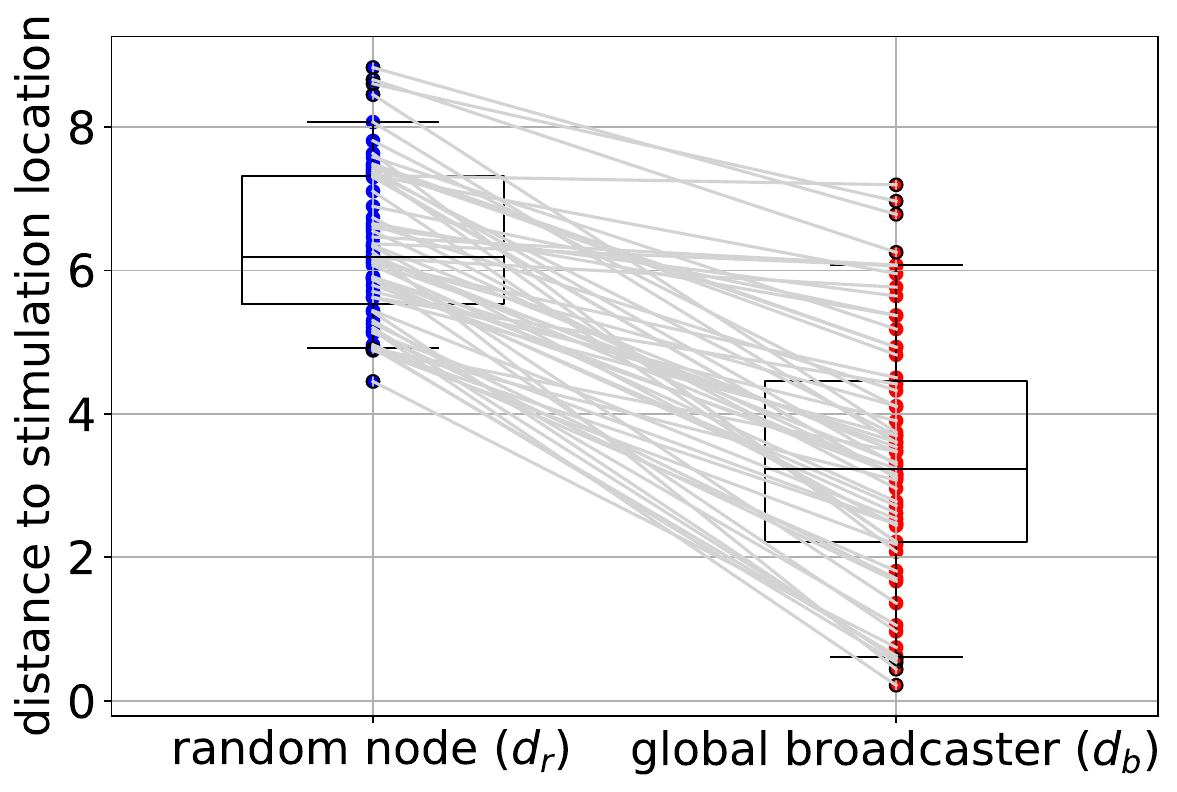}}

	\vspace{-0.2cm}
	\caption{Distance of global broadcaster from stimulation location $d_\mathrm{b}$ obtained from the gradient flow \SI{10}{\milli\second} after stimulation (right) compared to average distance of a random node to the stimulation location $d_\mathrm{r}$ for all 63 experimental sessions. A model order $K=5$ is used for estimating the flow. The box plots indicate $5^\mathrm{th}$ and $95^\mathrm{th}$ percentile, $1^\mathrm{st}$ and $3^\mathrm{rd}$ quartiles, and medians. The gray lines indicate the improvement for each individual session.}
	\label{fig:global_broadcaster}
	
\end{figure}

We computed $x_b$ from the gradient flow signal \SI{10}{\milli\second} after stimulation for all 63 experimental sessions as outlined above.
Then we computed the distance to the stimulation location of the fist laser
\begin{equation}\label{eq:dist_global_broadcaster}
	d_\mathrm{b} = \vert\vert x_\mathrm{b} - x_\mathrm{s} \vert\vert_2.
\end{equation}
This distance is shown in Fig.~\ref{fig:global_broadcaster} on the right.
On the left of Fig.~\ref{fig:global_broadcaster} we show the average distance of a random broadcaster to the stimulation location
\begin{equation}
	d_\mathrm{r} = \frac{1}{N} \sum_{n=1}^{N-1} \vert\vert l(n) - x_\mathrm{s} \vert\vert_2.
\end{equation}
One can clearly see that, as expected, the distance of the global broadcaster to the stimulation location $d_\mathrm{b}$ is significantly smaller than $d_\mathrm{r}$ (p-value: $3.1 \cdot 10^{-21}$; Student's t-test), which supports the hypothesis that the gradient flow depends on the stimulation location.

\section{Conclusion}\label{sec:conclusion}
We have illustrated how neural flow naturally arises as the result of a graph diffusion process.
Furthermore, this diffusion process can be extended to higher orders to allow for variable flow delays between different nodes of the graph.
To validate our diffusion model, we have applied it to neural recordings from monkeys obtained during excitatory optogenetics. 
Finally, we have demonstrated how the neural flow can be decomposed into a gradient and rotational component and that the gradient component depends on the location of stimulation showing flow spreading away from the stimulation site.

In the future, we plan to investigate the relationship between our diffusion model and vector autoregressive models that are widely used in the neuroscience community.
Additionally, we want to study the flow signals obtained from the diffusion model in more detail by further analyzing their spatiotemporal patterns and incorporating the rotational component.
With the advancement of electrode array technologies \cite{griggs_2019, griggs_2021}, theses novel processing techniques will become more essential in studying neural circuits and fine scale network dynamics.

\bibliographystyle{IEEEbib}
\bibliography{refs}

\end{document}